\documentclass[11pt]{article}

\newcommand{\etal}{{\it et al.\,\,}}
\newcommand{\brho}{\mbox{\boldmath $\rho$}}
\newcommand{\bOmega}{\mbox{\boldmath $\Omega$}}

\begin{document}

\title{\bf Quantum information processing: A linear-systems perspective} 

\author{Marcos Curty and David J.\ Santos\thanks{Marcos Curty and David J.
Santos are with the Departamento de Tecnolog\'{\i}as de las Comunicaciones,
Universidad de Vigo, Campus Universitario s/n. E-36200 Vigo (Spain).}}

\maketitle

\begin{abstract}
In this paper a system-oriented formalism of Quantum Information
Processing is presented. Its form resembles that of standard signal
processing, although further complexity is added in order to describe pure
quantum-mechanical effects and operations. Examples of the application
of the formalism to quantum time evolution and quantum measurement are given.
\end{abstract}

\section{Introduction}

The field of Quantum Information
Processing (QIP) has recently received a renewed
attention, particularly by the Physics and Computing communities
\cite{BENNETT_1995,DIVINCENZO_1995,PRESKILL_1999,STEANE_2000_A,STEANE_2000}.
This discipline, however, dates back from the sixties and seventies,
when the advent of the laser generated a lot of interest among
communication-theory researchers (see \cite{HELSTROM_1970,DAVIES_1977}
and references therein). This interest arose from the possibility
of employing the laser as an electromagnetic source in ultra-wideband
communication systems. In order to establish a suitable model of
the communication channel, and owing to the high frequency of the
laser radiation, it was necessary to consider in the modelling the
quantum-mechanical properties of radiation (quantum noise). These early
research efforts constituted the birth of Quantum Communication Theory. 

First explorations on the quantum nature of information are due to
MacKay \cite{MACKAY_1953}, although it was Stern \cite{STERN_1960} the
first, to our knowledge, who modelled optical communication channels
quantum-mechanically. Gordon \cite{GORDON_1962} also recognized the
influence of quantum noise in communication systems, and addressed the
problem of selecting the appropriate receiver structure for an optical
communication channel. Following the standard approach of Quantum Field
Theory, She \cite{SHE_1968} studied the channel and the receiver of a
communication system. Later, in what we consider was the most complete
research effort towards the full quantum-mechanical characterization
of the detection problem in an optical communication channel, Helstrom
\etal \cite{HELSTROM_1970} analyzed in detail how the laws of Quantum
Mechanics (QM) affect the reliability of a communication system, and
how to design quantum-mechanical receivers. These same issues were
also addressed in a similar manner by Liu \cite{LIU_1970}, Personick
\cite{PERSONICK_1971}, and Davies \cite{DAVIES_1977,DAVIES_1978}.

During the last twenty-five years work has continued on the
characterization of quantum receivers and its optimum design
\cite{HELSTROM_1976,HELSTROM_1979,SHAPIRO_1979,CHARBIT_1989,BENDJABALLAH_1998};
however, the full modelling of information-processing quantum-mechanical
systems with a broader scope (not limited to optical communication
channels) has only recently been proposed (see, e.g. \cite{BENNETT_1998}
and references therein), arguably fueled by the fertility in the field
of Quantum Optics during the eighties. As a consequence, brand-new
disciplines related to the quantum processing of information, such as
Quantum Computation, Quantum Cryptography and Quantum Teleportation, have
appeared \cite{STEANE_2000}. Behind these new disciplines we find again
QM, but unlike in the typical formulations of the sixties and seventies,
now the processing (in communication as well as computing) is described
making use of the theory of quantum open systems \cite{DAVIES_1976}.

QM is not an easy theory to grasp, and Open-Systems QM is even more
subtle, particularly to the Information-Processing community. With
this paper we expect to contribute to shorten the gap that separates
researchers with a Physics background from those more concentrated on
classical Signal Processing (SP). With this purpose in mind, we show
that a general formalism of QIP similar to that of classical SP can
be established. 

The paper is organized as follows. In Section~\ref{CLASICO}
we review the formalism of classical SP. In
Section~\ref{CUANTICO} we define the concept of quantum system and
discuss how to efficiently model the dynamics of such a system in general.
In Section~\ref{MATRICES} we derive the matrix representation of the
input-output relationship of a quantum system. 
Section~\ref{SISTEMAS_COMPUESTOS} generalizes the proposed formalism
to composite systems. This allows to describe the concept of
entanglement. In Section~\ref{EJEMPLOS} we apply the formalism to the
description of two key quantum processes: Quantum Evolution and
Quantum Measurement. Finally, in Section~\ref{CONCLUSIONES} we
summarize the main results of the paper.

\section{Classical signals and systems}
\label{CLASICO}

SP is concerned with the representation, transformation, and manipulation
of signals and the information they contain. Generally, this information
is about the state or behaviour of a physical system, although we
shall regard signals just as containers of information. Mathematically,
signals are represented as functions of one or more variables belonging
to a vector signal space.

Depending on the character of these variables, continuous or discrete,
SP has been divided into Analog and Discrete Signal Processing (DSP),
respectively.  Progress in computer technology has made possible
the easy implementation of the latter and its pervasiveness. Thus, in
the following we shall restrict ourselves to DSP.  

Discrete signals are represented by sequences of numbers that we shall
denote, following the standard formalism of one-dimensional DSP, by
$x[n]$ ($n=1,\cdots,N$). The actual processing of signals is performed
by systems, that can be seen as entities that map an input signal
$x[n]$ to an output signal $y[n]$. A particularly important class of
systems  is formed by those that are linear and time-invariant. These
two properties  allow to write the system input-output relationship by
means of the convolution operator:

\begin{equation}
y[n]=x[n]*h[n]=\sum_{k=1}^N x[k]h[n-k],
\label{ENTRADA_SALIDA_CLASICA}
\end{equation}

\noindent where the sequence $h[n]$ is the impulse response of the
system. The relationship expressed by equation
(\ref{ENTRADA_SALIDA_CLASICA}) can also be written in the transformed
domain by means of the Discrete Fourier Transform (DFT):

\begin{equation}
Y[k]=X[k]H[k],
\label{ENTRADA_SALIDA_CLASICA_DFT}
\end{equation}

\noindent where the capitalized sequences are the DFTs of the
corresponding signals.

Besides linearity and time-invariance, other restrictions
of more fundamental physical nature, such as stability and causality,
can be imposed on the operation of systems. Further information regarding
these and similar issues can be seen in \cite{OPPENHEIM_1989}.

\section{Quantum signals and systems}
\label{CUANTICO}

Just like a classical system, a quantum system is a mathematical entity that 
operates on the
quantum signal at its input and generates an output quantum signal.
Quantum signals are often called states, and, following the standard
QM formalism, are represented by vectors belonging to a Hilbert signal
space ${\cal E}$. In ``bra-ket'' or Dirac notation, we shall write the input and
output signals of a system as $|x\rangle$ and $|y\rangle$, respectively.
We shall assume in the following that the signal space has a finite
dimension $N$. This is a requirement in order to describe discrete 
quantum signals.

Quantum systems, as classical ones, can manipulate two types of signals,
pure and mixed (stochastic).  Pure signals are deterministic; on the
contrary, mixed signals result from the statistical mixture of pure
ones. As an example, consider the case in which the input signal can be
one of the set $\{|x_i\rangle; \, \, i=1,\cdots,M\}$ with probability
$p_i$. In  situations like this, the information about the input of the
system can be better described by an hermitian operator called density
operator (see \cite{FANO_1957} for a definition of density operators and
their more relevant properties) that we write as:

\begin{equation}
\hat \rho_x=\sum_{i=1}^M p_i |x_i\rangle\langle x_i|.
\label{OP_X}
\end{equation}

Pure signals can also be represented in terms of density operators. If
the input signal $|x\rangle$ is deterministic (has probability one), then

\begin{equation}
\hat \rho_x=|x\rangle\langle x|.
\end{equation}

\noindent Since this formalism allows to describe both types of quantum
signals, in the following we shall characterize the signals at the input
and output of our systems by the density operators $\hat \rho_x$ and
$\hat \rho_y$, respectively. As for the relation between the input and
output signals, it can be expressed by means of a superoperator $\$$:

\begin{equation}
\hat \rho_y=\$(\hat \rho_x).
\end{equation}

\noindent Superoperators are complete positive operators that transform
density operators preserving its hermiticity and trace properties (see
\cite{KRAUS_1983} for an account of superoperator theory). As we shall
study linear quantum systems, superoperators will be also necessarily
linear. 

In its more general form, the action of a superoperator on a quantum
signal can be written as 

\begin{equation}
\hat \rho_y=\sum_\mu \hat M_\mu \hat \rho_x \hat M_\mu^\dagger,
\label{REP_KRAUS}
\end{equation}

\noindent where the $\hat M_\mu$ operators, called Kraus operators, satisfy the
closure relation 

\begin{equation}
\sum_\mu \hat M_\mu^\dagger \hat M_\mu=\hat 1.
\label{CONDICION_MUS}
\end{equation}

\noindent It can be shown \cite{KRAUS_1983} that the necessary number of
Kraus operators is at most $N^2$. Equation (\ref{REP_KRAUS}) is the
Kraus representation of the density operator $\hat \rho_y$. When this
representation requires only one Kraus operator, the system is said
to be invertible since the Kraus representation reduces to a unitary
transformation. It is remarkable, and we shall have the opportunity to
give examples below, that any conceivable quantum system can be modelled
by a superoperator.

\section{Matrix representation of QIP}
\label{MATRICES}

In the previous section we have seen that quantum signals are vectors
belonging to a signal space and, more generally, density operators. Up
to now the discussion has remained rather abstract. In this section we
shall develop a matrix formalism of the input-output theory previously
presented.

Consider an arbitrary pure signal $|x\rangle$ and let us define on ${\cal
E}$ the orthonormal base $\{ |u_i\rangle;\,\, i=1,\dots,N\}$. In this
base $|x\rangle$ can be written as:

\begin{equation}
|x\rangle=\sum_{n=1}^N x[n] |u_n\rangle,
\end{equation}

\noindent where the coefficients of the expansion are given by the
inner products $\langle u_n|x\rangle$. Given the base used,
these coefficients univocally determine the information contained in
the signal. Therefore, in a given base, the quantum signal can be
represented by the sequence $x[n]$, much as in classical DSP.

Let us turn now to the more general case described by density operators. 
When the signal-space base just introduced is used on the signals
$|x_i\rangle$ of Section~\ref{CUANTICO}, we get

\begin{equation}
|x_i\rangle=\sum_{n=1}^N x_i[n] |u_n\rangle,
\end{equation}

\noindent where $x_i[n]=\langle u_n|x_i\rangle$. If this expansion is
used in equation (\ref{OP_X}), the following representation of the
input density operator is obtained:

\begin{equation}
\hat \rho_x=\sum_{n,m=1}^N \rho_x[n,m] |u_n\rangle\langle u_m|,
\end{equation}

\noindent where 

\begin{equation}
\rho_x[n,m]=\sum_{i= 1}^M p_i x_i[n]x_i^*[m].
\label{ELEMENTOS}
\end{equation}

\noindent Now the information at the input of the system is
contained in the hermitian matrix $\brho_x$ whose elements are given by
equation (\ref{ELEMENTOS}). From the properties
of density operators \cite{FANO_1957} it can be demonstrated that: 

\begin{enumerate}
\item The eigenvalues of $\brho_x$ are always between 0 and 1.
\item The trace of $\brho_x$ is 1.
\item $\brho_x$ is positive definite.
\end{enumerate}

Note that in the case of the pure signal $|x\rangle$,
$\brho_x$ reduces to an idempotent matrix of elements 
$\rho_x[n,m]=x[n]x^*[m]$.

The action of a superoperator can also be written in matrix form 
once a base of the signal space is chosen. Using the same base as above, the
Kraus representation of the quantum output signal, see equation 
(\ref{REP_KRAUS}), is given by

\begin{equation}
\rho_y[n,m]=\sum_\mu \sum_{i,j=1}^NM_\mu[n,i]\rho_x[i,j]M_\mu^\dagger[j,m],
\label{ENTRADA_SALIDA_CUANTICA}
\end{equation}

\noindent where $M_\mu[k,l]=\langle u_k|\hat M_\mu|u_l\rangle$ and,
from (\ref{CONDICION_MUS}),

\begin{equation}
\sum_{\mu} \sum_{k=1}^NM_\mu^\dagger[i,k] M_\mu[k,j]=\delta_{ij}.
\label{CONDICION_MUS_MATRICIAL}
\end{equation}

\noindent Equation (\ref{ENTRADA_SALIDA_CUANTICA}) can be regarded as
the equivalent of the classical input-output relationships
(\ref{ENTRADA_SALIDA_CLASICA}) or 
(\ref{ENTRADA_SALIDA_CLASICA_DFT}) in QIP. In general, the dynamics of
quantum systems can be described employing just standard matrix algebra:
the output signal is obtained as the sum of all the congruence matrix
transformations of the input signal given by the matrices associated
to the Kraus operators.  In compact form:

\begin{equation}
\brho_y=\sum_\mu {\bf M}_\mu \brho_x {\bf M}_\mu^\dagger.
\label{FORMA_COMPACTA}
\end{equation}

\noindent Notice the more complex character of the quantum formalism as
compared to the classical one. This complexity comes from the fact
that the quantum description has to incorporate the possibility of
quantum decoherence, i.e. the transformation of an input
pure signal into a mixed output one.

In order to clarify the concepts introduced, we present a simple
example. Consider the quantum-mechanical binary communication
channel depicted in Figure~1. The signal space in this case is
two-dimensional. If we denote the two elements of a base of this space by
$|0\rangle$ and $|1\rangle$, then any signal (ordinarily called qubit,
quantum bit) can be written as the superposition 
$\alpha |0\rangle+\beta |1\rangle$,
where $|\alpha|^2+|\beta|^2=1$. Let us assume that the input signal can
be $|0\rangle$ or $|1\rangle$. According to the figure, the output
signal can also be $|0\rangle$ or $|1\rangle$ with the probabilities
shown in the figure. In this simple case the matrix form of one set 
of Kraus operators that describes the channel is:

\begin{equation}
{\bf M}_1=\sqrt{p}\left(
\begin{array}{cc}
1 & 0\\
0 & 1
\end{array}
\right),\,
{\bf M}_2=\sqrt{1-p}\left(
\begin{array}{cc}
0 & 1\\
0 & 0
\end{array}
\right),\,
{\bf M}_3=\sqrt{1-p}\left(
\begin{array}{cc}
0 & 0\\
1 & 0
\end{array}
\right).
\end{equation}

These Kraus operators can also be used to obtain the output signal when
at the input of the channel we have a signal of the more general form
$\alpha |0\rangle+\beta |1\rangle$:

\begin{equation}
\brho_y=\left(
\begin{array}{cc}
p|\alpha^2|+(1-p)|\beta|^2 & p\alpha\beta^*\\
p\alpha^*\beta & p|\beta^2|+(1-p)|\alpha|^2 
\end{array}
\right).
\end{equation}

\noindent In this case the input signal will remain the same with
probability $p$, will transform to $|0\rangle$ with probability
$(1-p)|\alpha^2|$, and to $|1\rangle$ with probability $(1-p)|\beta^2|$.

\section{Composite quantum systems: Entangled signals}
\label{SISTEMAS_COMPUESTOS}

Many of the proposed applications of QIP, such as Quantum Computation,
Quantum Cryptography and Quantum Communications are based on the
concept of information entanglement (see, e.g. \cite{WILLIAMS_1998}).
Entangling information in two or more signals has some advantages over
the one-signal-one-system QIP introduced in the previous section. In this
section we present composite systems. We shall show that the input-output
formalism previously presented can also describe this kind of quantum
systems.

Consider two quantum systems $A$ and $B$ whose input and output signals
are, respectively, the density operators $\hat \rho_x^A$, $\hat
\rho_x^B$, $\hat \rho_y^A$, and $\hat \rho_y^B$. Let us assume that
the input-output relationship for each system is known (in the form
of two superoperators $\$_A$ and $\$_B$), and that there is some
interaction between the actual physical processes modelled by the two
systems. The nature of this interaction is not relevant to our current
purpose. One can define a global or composite system that comprises
the two systems and its physical interaction, and global input and output
signals. In standard QM this is described
making use of the tensor product of signal spaces: If ${\cal E}_A$
and ${\cal E}_B$ are, respectively, the signal spaces in systems $A$
and $B$, then ${\cal E}_A\otimes {\cal E}_B$ is the signal space of
the global signals. Therefore, the composite system $AB$ has global
input and output signals given, respectively, by $\hat \rho_x^{AB}$
and $\hat \rho_y^{AB}$, and its input-output relation is described by
some global superoperator:

\begin{equation}
 \hat \rho_y^{AB}=\$_{AB}(\hat \rho_x^{AB}). 
\end{equation}

Like the signals and systems of the previous section, composite systems
and global signals have also a matrix representation. If the bases of
${\cal E}_A$ and ${\cal E}_B$ are, respectively,  $\{|a_n\rangle;\,\,
n=1,\cdots,N\}$ and $\{|b_n\rangle;\,\, n=1,\cdots,N\}$ (we are assuming,
for simplicity, that both signal spaces have the same dimension, $N$),
the base of ${\cal E}_{AB}$ is $\{|a_n\rangle\otimes |b_m\rangle;\,\,
n,m=1,\cdots,N\}$. In this base the operator density associated to the
global input signal is

\begin{equation}
\hat \rho_x^{AB}=\sum_{n,m,i,j=1}^N\rho_x^{AB}[n,m;i,j]
|a_n,b_m\rangle\langle a_i,b_j|.
\end{equation}

\noindent As the number of signals and systems is doubled, so
does the complexity of the description, and now the signals are
described by rank-four tensors. Notice, however, that the information
contained in the tensor can be arranged in a standard square matrix with elements
$\rho_x^{AB}[k,l]$, where the indexes $k$ and $l$ ($k,l=1,\cdots,N^2$)
enumerate, respectively, the different ($N^2$ in each case) pairs $(n,m)$
and $(i,j)$.

Although we have described the action of the composite system globally, in some
situations it may be necessary to access the $A$ or $B$ part of
a global signal.  This can be accomplished using the partial trace
\cite{FANO_1957}:

\begin{eqnarray}
      & &\rho_y^A[n,i]={\rm Tr}_B\{\rho_y^{AB}[n,m;i,j]\}=   
\sum_{m=1}^N\rho_{y}^{AB}[n,m;i,m],\\
      & &\rho_y^B[m,j]={\rm Tr}_A\{\rho_y^{AB}[n,m;i,j]\}=   
\sum_{n=1}^N\rho_{y}^{AB}[n,m;n,j].
\end{eqnarray}

\noindent This may be the case of a physical system that interacts undesirably
with another. Using the equations above the influence of
the disturbing system on the dynamics of the main system can be studied.

In some circunstances, global signals of composite systems cannot be
factored accordingly to the tensor product defined on the global signal
space, i.e. $\hat \rho^{AB}\neq\hat \rho^A\otimes\hat \rho^B$, or,
more generally, $\hat \rho^{AB}\neq\sum_i p_i \hat \rho_i^A\otimes\hat
\rho_i^B$. This means that the information contained in all the partial
signals is not equivalent to the information contained in the global
signal. Thus, quantum information, unlike its classical counterpart, can
be encoded in nonlocal correlations between the different parts of a
composite system. When this happens, signals are said to be entangled. 

In order to clarify the ideas presented in this section,
assume that the signal at the input of system $A$ is
$|\psi_x\rangle_A=\alpha|0\rangle_A+\beta|1\rangle_B$ and that the signal
at the input of system $B$ is $|\psi_x\rangle_B=|0\rangle_B$. In the
bases $\{|0\rangle_A,|1\rangle_A\}$ and $\{|0\rangle_B,|1\rangle_B\}$,
the input signals can be written, respectively, as

\begin{equation}
\brho_x^A=\left(
\begin{array}{cc}
|\alpha^2| & \alpha\beta^*\\
\alpha^*\beta & |\beta^2| 
\end{array}
\right),
\end{equation}

\noindent and

\begin{equation}
\brho_x^B=\left(
\begin{array}{cc}
1& 0\\
0& 0 
\end{array}
\right).
\end{equation}

\noindent In the natural base of the composite system,
$\{|00\rangle,|01\rangle,|10\rangle,|11\rangle\}$, the global input signal is
represented by the matrix

\begin{equation}
\brho_x^{AB}=\left(
\begin{array}{cccc}
|\alpha^2|&0&\alpha\beta^*&0\\
0&0&0&0\\
\beta\alpha^*&0&|\beta|^2 &0\\
0&0&0&0
\end{array}
\right).
\label{ENTRADA_GLOBAL}
\end{equation}

\noindent Consider now that the composite system $AB$ produces  the
unitary  tranformation  $\brho_y^{AB}={\bf M}\brho_x^{AB}{\bf
M}^\dagger$
on the global input signal, where

\begin{equation}
{\bf M}=\left(
\begin{array}{cccc}
1&0&0&0\\
0&1&0&0\\
0&0&0&1\\
0&0&1&0
\end{array}
\right).
\end{equation}

\noindent For the input given by equation 
(\ref{ENTRADA_GLOBAL}), the corresponding global output signal is

\begin{equation}
\brho_y^{AB}=\left(
\begin{array}{cccc}
|\alpha|^2&0&0&\alpha\beta^*\\
0&0&0&0\\
0&0&0&0\\
\alpha^*\beta&0&0&|\beta|^2
\end{array}
\right).
\end{equation}

\noindent It can be shown that, contrary to what happened at the input,
at the output the signal cannot be expressed as the tensor product of a
signal from ${\cal E}_A$ and another from ${\cal E}_B$. The signals, 
that were uncorrelated at the input, are now entangled. 

\section{Some quantum systems}
\label{EJEMPLOS}

In Section \ref{MATRICES} we have shown that the operation of any quantum
system can be described in matrix form (see equation
(\ref{FORMA_COMPACTA})). In this section we are going to show that the
matrix formalism proposed can in fact be applied to the description of
two typically quantum processes: quantum evolution, and quantum
measurement.

\subsection{Time evolution described as a system operation}
\label{TIEMPO}

Physically, any kind of processing requires a lapse of time. In the
systems previously considered we have assumed that processes occur
instantaneously, so no time dependence has been included in the
formalism. Consider now the case of a quantum system whose input is
given by $\brho_x$ and its output, the processed input, by 
$\brho_y$. If the processing begins at the instant $t_0$ and finishes at
$t$, the laws of QM say (see, e.g. \cite{COHEN_1977}) that both quantum
signals are related by the unitary transformation

\begin{equation}
\brho_y= {\bf U}(\Delta \tau) \brho_x {\bf U}^\dagger(\Delta \tau),
\label{EVOLUCION_MATRICIAL}
\end{equation}

\noindent where ${\bf U}$ is the matrix representation of the standard
evolution operator $\hat U =\exp(-i \hat H \Delta \tau /\hbar)$, with
$\hat H$ the Hamiltonian of the system, and $\Delta \tau=t-t_0$.
Equation (\ref{EVOLUCION_MATRICIAL}), owing to the unitarity of the
matrix ${\bf U}$, is a particular case of equation
(\ref{FORMA_COMPACTA}), and thus can be seen as a Kraus representation.

The evolution represented by equation (\ref{EVOLUCION_MATRICIAL}) is
characteristic of isolated quantum processes, i.e. closed physical
systems. However, in some cases two or more processes are coupled
forming a composite physical system and one is interested in the time
evolution of the global signal and its different parts. This composite
physical system can be regarded as a closed system, but the parts (the
so-called open systems) certainly not. This situation, as we show below,
can also be described by the matrix formalism.

As in Section \ref{SISTEMAS_COMPUESTOS}, assume we have two interacting
physical systems, $A$ and $B$, that unite to form the composite physical
system $AB$. This system is closed, so it can be described by an
equation like (\ref{EVOLUCION_MATRICIAL}):

\begin{equation}
\brho_y^{AB}= {\bf U}_{AB}(\Delta \tau) \brho_x^{AB} 
{\bf U}_{AB}^\dagger(\Delta \tau),
\label{EVOL_MATRICIAL}
\end{equation}

\noindent where the Hamiltonian inside ${\bf U}$ takes into account the
dynamics of both physical systems as well as their mutual interaction. Taking the
partial trace, the matrix representation of the time-evolution of one of
the partial systems, say $A$, can be easily obtained: $\brho_y^A={\rm
Tr}_B\{\brho_y^{AB}\}$. However, it is slightly more involved to obtain
the Kraus representation of this input-output relationship. If the
two systems are initially uncorrelated, i.e. $\hat \rho_x^{AB}=\hat
\rho_x^A\otimes\hat \rho_x^B$, it can be shown \cite{SCHUMACHER_1996}
that 

\begin{equation}
\brho_y^A=\sum_{n=1}^N \bOmega_n \brho_x^A \bOmega_n^\dagger,
\label{SIN_DEMOSTRAR}
\end{equation}

\noindent where the $\bOmega_n$ matrices, that obey the condition 
(\ref{CONDICION_MUS_MATRICIAL}), are function of the evolution of
the composite system, ${\bf U}_{AB}(\Delta \tau)$, and of $\brho_x^B$.

\subsection{Measurement described as a system operation}

The quantum description of measurement is still rather controversial,
particularly among researchers focused on the interpretative
aspects of QM \cite{FUCHS_2000}. Measurement can be described
ideally as a process that takes place instantaneously (the von Neumann
or othogonal measurement) in a closed system, or taking into account
that any measurement needs an interaction, and that interactions do not
occur (at least in the physical world) instantaneously (the generalized
measurement in an open system) \cite{LUIS_1999}. We shall show that 
the general formalism developed in Section \ref{MATRICES} can describe
both kinds of measurement processes.

Consider a quantum system that performs a measurement on its input
signal $\brho_x$ to produce at its output, as a result of such a
measurement, the signal $\brho_y$. Associated to any measurement there
is an observable $\hat Q$ (an hermitian operator; see, e.g. \cite{COHEN_1977})
with the eigenvalue problem (for notational simplicity we study the
non-degenerate case)

\begin{equation}
\hat Q |q_n\rangle = q_n |q_n\rangle.
\end{equation}

\noindent The eigenstates of this problem are a base of the signal
space, and the family of orthogonal projectors $\{|q_n\rangle\langle
q_n|; n=1,\cdots,N\}$ are such that $\sum_{n=1}^N|q_n\rangle\langle
q_n|=\hat 1$. It can be shown from QM \cite{COHEN_1977} that, in operator
language, 

\begin{equation}
\hat \rho_y = \sum_{n=1}^N (|q_n\rangle \langle q_n|)\hat \rho_x
(|q_n\rangle\langle q_n|).
\label{RISA}
\end{equation}

\noindent This equation can be written in matrix form once a base of the
signal space is selected:

\begin{equation} 
\brho_y=\sum_{n=1}^N {\bf M}_n \brho_x {\bf M}^\dagger_n,
\end{equation} 

\noindent where the elements of the Kraus matrices are given by $
M_n[i,j]=\langle a_i|q_n\rangle\langle q_n|a_j\rangle$, and we have
used the base $\{|a_n\rangle, n=1,\cdots,N\}$.

As we mentioned at the beginning of this subsection, any physical
measurement requires the interaction of the signal with (at least) 
the measuring apparatus, and this interaction is not an instantaneous
physical process. The quantum-mechanical consideration of this fact
leads to a more complex description of measurement based on Positive
Operator Valued Measures (POVM) \cite{BRANDT_1999}. In these generalized
measurements the signal to be measured interacts with another called
ancilla and an orthogonal measurement is performed on the ancilla part
of the resulting global signal. Owing to the interrelation between
both signals, this measurement gives information about the original
signal.

Suppose we have a signal $\hat \rho_x^A$ that interacts with another
signal another initially uncorrelated signal $\hat \rho_x^B$. The time
evolution of this interaction, according to what we saw in
Subsection~\ref{TIEMPO}, can be modelled by the unitary Kraus
representation (\ref{EVOL_MATRICIAL}). Let us now perform the orthogonal
measurement of $\hat Q$ on the $B$ part of the output
signal $\brho_y^{AB}$, and leave the $A$ part untouched.  According to
the previous formalism for orthogonal measurements, see equation 
(\ref{RISA}), the resulting composite signal is

\begin{equation}
\hat \rho_z^{AB}=\sum_{k=1}^N
(|q_k\rangle\langle q_k|\otimes\hat 1^A) 
\hat U_{AB}(\Delta \tau) \hat \rho_x^{AB} U_{AB}^\dagger(\Delta \tau)
(|q_k\rangle\langle q_k|\otimes\hat 1^A).
\end{equation}

\noindent If we now take the partial trace over $B$, the output signal of system $A$
after the generalized measurement can be written as

\begin{equation}
\brho_z^A={\rm Tr}_B\{{\bf U}_{AB}\brho_x^{AB}{\bf U}_{AB}^\dagger\}.
\end{equation}

\noindent An input-output
relationship of the form of equation (\ref{SIN_DEMOSTRAR}) can be
obtained for the $A$ system, just as we did in Section~\ref{TIEMPO}.

\section{Conclusion}
\label{CONCLUSIONES}

We have shown that any quantum-mechanical information processing can
be seen as the operation of an abstract system with input and output
signals. Both signals and systems are represented by matrices. The
output signal is obtained as a superposition of congruence matrix
transformations on the input signal. Our main objective in proposing
such a formalism is to provide a framework to further study QIP suitable
for non-physicists. As two applications of the formalism, we have
expressed in signal-system form the processes of quantum time-evolution
and quantum measurement.


\end{document}